\documentclass{article}
\usepackage[margin=1.25in]{geometry}
\usepackage{graphicx}
\usepackage{amsmath}
\usepackage{amssymb}
\usepackage{parskip}
\usepackage[hidelinks]{hyperref}
\usepackage{graphicx}
\usepackage{natbib}
\usepackage[bf]{caption}
\usepackage{subcaption}

\usepackage{comment}
\usepackage{color}
\begin{document}
\title{
	On modelling bicycle power-meter measurements:~Part~I \\
	{\Large Estimating effects of air, rolling and drivetrain resistance}
}
\author{
	Tomasz Danek%
	\footnote{
		AGH--University of Science and Technology, Krak\'{o}w, Poland, \texttt{tdanek@agh.edu.pl}
	}\,,
	Michael A. Slawinski%
	\footnote{
		Memorial University of Newfoundland, Canada, \texttt{mslawins@mac.com}
	}\,,
	Theodore Stanoev%
	\footnote{
		Memorial University of Newfoundland, Canada, \texttt{theodore.stanoev@gmail.com}
	}
}
\date{May 9, 2020}
\maketitle
\begin{abstract}
Power-meter measurements together with GPS measurements are used to study the model that accounts for the use of power by a cyclist.
The focus is on estimating the coefficients of the air, rolling and drivetrain resistance, uncertainties of these estimates, as well as relations between them.
Expressions used in the main text are derived in the appendices.
\end{abstract}
\section{Introduction}
Using power meters to study cycling performance combines two distinct realms that do not have an explicit relation between them: mechanical measurements and physical conditions.
A hypothetical relation between them is offered by mathematical modelling.
For instance, the answer to a question\,---\,is a high power output that results in a low ground speed a consequence of a strong headwind, steep climb, their combination or a completely different factor, such as an unpaved road?\,---\,cannot be obtained from power meters alone; it can only be postulated, under various assumptions, as a model, without a claim to uniqueness of solution.

Many studies examine the physics of cycling.
For instance, there are wind-tunnel studies to measure the aerodynamics of bicycle wheels~\citep[e.g.,][]{GreenwellEtAl1995}, and the studies to estimate the accuracy of power measurements based on the frequency of the pedal-speed measurements~\citep[e.g.,][]{Favero2018}.
There are studies to examine power required to overcome the effect of winds, taking into account tire pressure, wheel radius, altitude and relative humidity~\citep[e.g.,][]{OldsEtAl1995}, as well as the aerodynamic drag, rolling resistance, and friction in the drivetrain~\citep[e.g.,][]{MartinEtAl1998}.
There are studies to estimate model parameters from measurements on the road~\citep[e.g.,][]{Chung2012} and to devise optimal speeds for time trials on closed circuits in the presence of wind~\citep[e.g.,][]{Anton2013}.
There are studies to investigate the aerodynamics of track cycling to predict the individual-pursuits times~\citep[e.g.,][]{UnderwoodJermy2014} and to simulate cyclist slip and steer angles necessary to navigate turns on a banked track~\citep[e.g.,][]{FittonSymons2018}.
There are graduate theses in mechanical engineering~\citep[e.g.,][]{Moore2008,Underwood2012}.

Be that as it may, the science of cycling is a rich field that combines theoretical, computational and experimental aspects of such disciplines as mathematical physics, continuum mechanics, as well as the optimization and approximation theories.

In this article, we consider power-meter measurements and a mathematical model to examine the conversion of power generated by a cyclist into motion of a bicycle.
In particular, we infer the values of the air, rolling and drivetrain resistance by seeking an acceptable agreement between obtained measurements and model retrodictions.
To do so, we combine classic formulations of fluid mechanics with innovative optimization methods, which extend the work of~\citet{Cavazzuti2012}.
We also invoke aspects of approximation theory to comment on the empirical adequacy of estimated values.

We begin this article by presenting the power-meter and GPS measurements.
Subsequently, we formulate and discuss a mathematical model to connect these two types of measurements.
Using this model and data from a flat segment of several kilometres\,---\,in Northwestern Italy, between Rivalta Bormida and Pontechino\,---\,that did not require any braking, we estimate the effects of the air, rolling and drivetrain resistance, as well as examine their uncertainties.
We conclude by a discussion of results.

This article contains also several appendices, where we present the derivations of expressions used in the main text to emphasize their assumptions and, hence, limitations.
These derivations are familiar to mathematical physicists, but might be less so to a broad range of sport scientists; as such, these appendices might be viewed as a brief auxiliary tutorial to facilitate the understanding of the main text. 
\section{Formulation}
\subsection{Power-meter measurements}
Power is a rate at which work is done; hence, it is equal to the amount of work divided by the time it takes to do it, which is tantamount to the product of force and speed,
\begin{equation}
	\label{eq:formula}
	P
	=
	f_{\circlearrowright} v_{\circlearrowright}
	\,.
\end{equation}
In the context of cycling, $f_{\circlearrowright}$ is the force applied to pedals and $v_{\circlearrowright}$ is the speed with which the rotating pedals cover the distance along the circumference of their rotation, which means that $v_{\circlearrowright}$ is proportional to the length of the crank.

\begin{figure}
	\centering
	\includegraphics[scale=0.7]{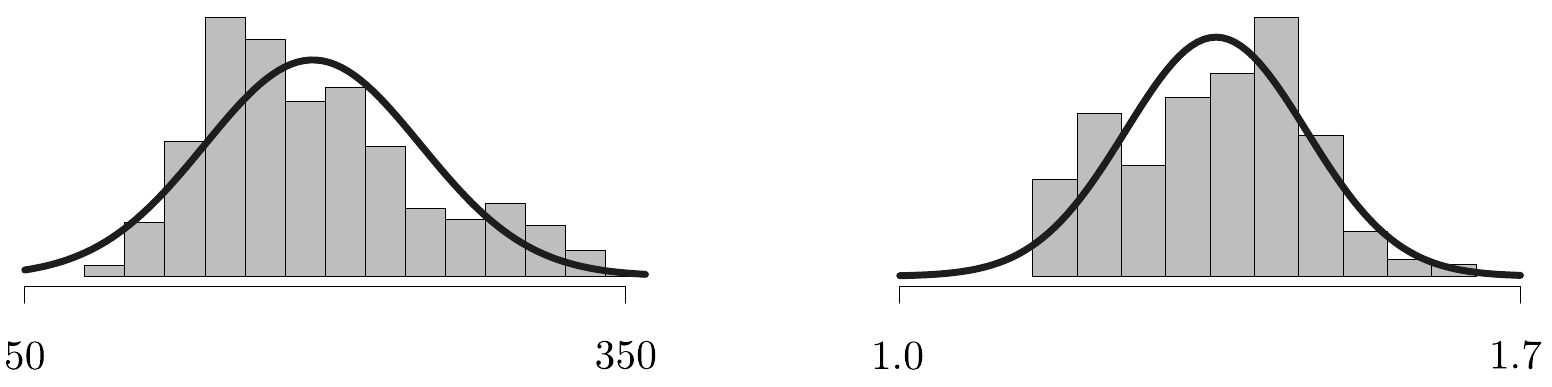}
	\caption{\small 
		Left-hand plot: force applied to pedals, $\bar f_{\circlearrowright}=193.8\pm 54.05$\,; right-hand plot: circumferential speed of pedals, $\bar v_{\circlearrowright}=1.357\pm 0.10226$
	}
	\label{fig:FigPedalVFFull}
\end{figure}

\begin{figure}
	\centering
	\includegraphics[scale=0.7]{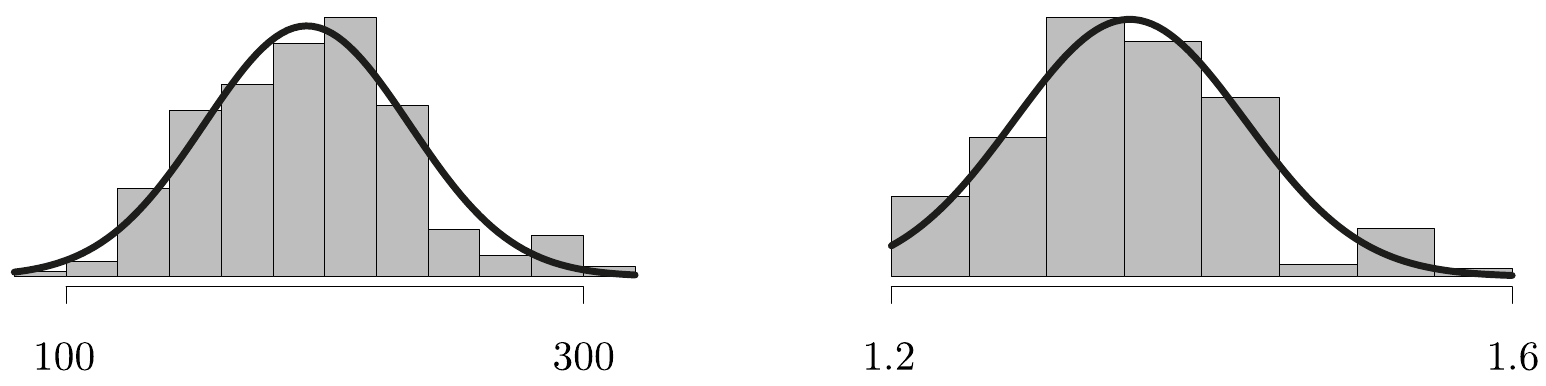}
	\caption{\small 
		Left-hand plot: force applied to pedals, $\bar f_{\circlearrowright}=193.1\pm 39.75$\,; right-hand plot: circumferential speed of pedals, $\bar v_{\circlearrowright}=1.354\pm 0.07448$
	}
	\label{fig:FigPedalVF}
\end{figure}

In this study\,---\,with the method used by \citet{Favero2018}\,---\,$v_{\circlearrowright}$ is an instantaneous speed, not an average per revolution; so is $f_{\circlearrowright}$\,.
The importance of such an approach is illustrated in Figures~\ref{fig:FigPedalVFFull} and \ref{fig:FigPedalVF},%
\footnote{For consistency with power meters, whose measurements are expressed in watts, which are $\rm{kg\,m^2/s^3}$\,, we use the {\it SI} units for all quantities.
Mass is given in kilograms,~$\rm{kg}$\,, length in meters,~$\rm{m}$\,, and time in seconds,~$\rm{s}$\,; hence, speed is in $\rm{m/s}$\,, change in speed in $\rm{m/s^2}$\,, and force in newtons,~$\rm{kg\,m/s^2}$\,; angle is in radians.}
which present the averages over pedal revolutions for the entire course, not the averages over a single pedal revolution or over a specific period; these averages discussed in Appendix~\ref{sec:AppendixA}.

\begin{figure}
	\centering
	\includegraphics[scale=0.7]{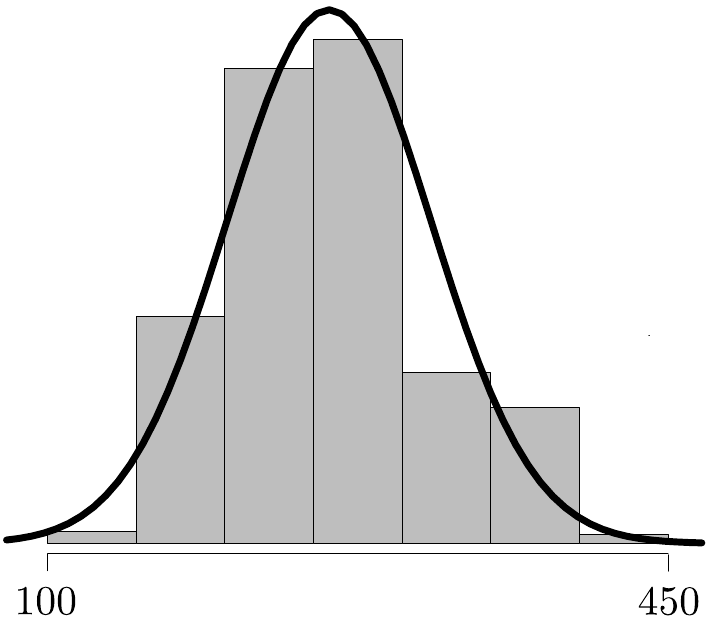} 
	\caption{\small
		Power: $\overline{P}=258.8\pm57.3$
	}
	\label{fig:FigPower}
\end{figure}

For measurements presented in Figure~\ref{fig:FigPedalVFFull}, the covariance between the two quantities is ${\rm cov}(v_{\circlearrowright},f_{\circlearrowright})=-4.1285$\,.
The average of the product is $\overline{v_\circlearrowright f_\circlearrowright}=258.811$\,, which is consistent with $\overline{P}=258.8$\,, stated in the caption of Figure~\ref{fig:FigPower}, and based on power-meter measurements. 
This is distinct from the product of the averages, $\overline v_\circlearrowright\,\overline f_\circlearrowright=262.9321$\,.
The two are related by the covariance, $\overline v_\circlearrowright\,\overline f_\circlearrowright+{\rm cov}(v_\circlearrowright,f_\circlearrowright)=262.9321-4.128505=258.8036$\,.

Measurements presented in Figure~\ref{fig:FigPedalVF}, in contrast to Figure~\ref{fig:FigPedalVFFull}, contain only the values between first and third quartile of recorded speeds.
In general, this restriction eliminates the data associated with the lowest and highest speeds, as could be the case of deceleration prior to, and acceleration following, a turnaround, as well as other random events or spurious data.
In this study, the differences between the average of the product and the product of the averages, and between Figures~\ref{fig:FigPedalVFFull} and \ref{fig:FigPedalVF} are small, since they correspond to a steady ride beginning with a flying start.
In other cases, such as a sprint from a standing start, they can be significant.
The power\,---\,which, according to expression~(\ref{eq:formula}), is the product of $v_{\circlearrowright}$ and $f_{\circlearrowright}$\,---\,is illustrated in Figure~\ref{fig:FigPower}.
\subsection{GPS measurements}
\begin{figure}
	\centering
	\includegraphics[scale=0.7]{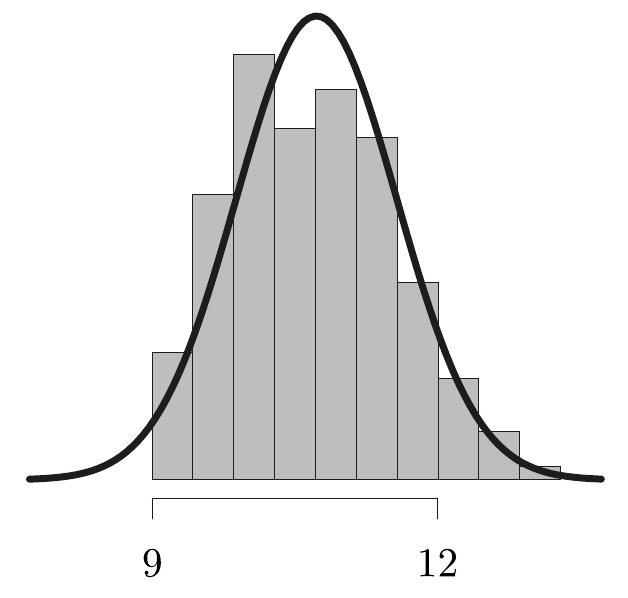}
	\caption{\small
		Ground speed, $\overline V_{\!\rightarrow}=10.51\pm0.9816$
	}
	\label{fig:FigVel}
\end{figure}

\begin{figure}
	\centering
	\includegraphics[scale=0.55]{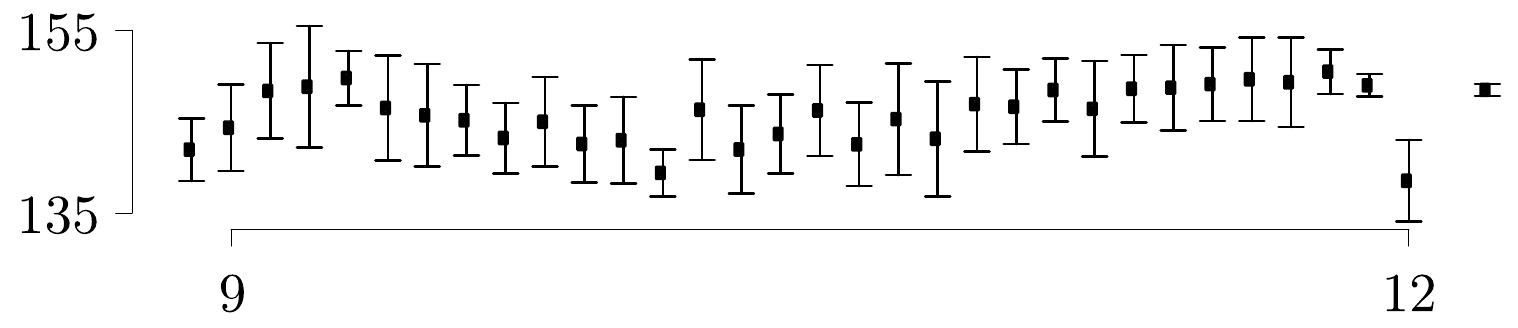}
	\caption{\small 
		Altitude}
	\label{fig:FigAltFlat}
\end{figure}

At the same time as the power-meter information is collected, the GPS collects information about speed, whose average values over the entire segment are illustrated in Figure~\ref{fig:FigVel}, and altitude, illustrated in Figure~\ref{fig:FigAltFlat}, where its values are grouped within speed intervals and the corresponding standard deviations are illustrated by error bars.
The average altitude over the entire segment is~$\overline{h} = 145.4\pm5.355$\,, and the median is $146.154\approx\overline{h}$\,, which is indicative of little change of altitude.

To relate the measured power to the surrounding conditions, we mediate between the two sources of information by a mathematical model.
Herein, the power meter provides information regardless of the conditions, GPS provides information about the surroundings independently of the power output, and a mathematical model\,---\,based on physical principles\,---\,serves as a hypothetical relation between them.
It must remain hypothetical since there is no explicit relation between the two sources of information.
\subsection{Mathematical model}
\begin{figure}
\centering
\includegraphics[scale=0.4]{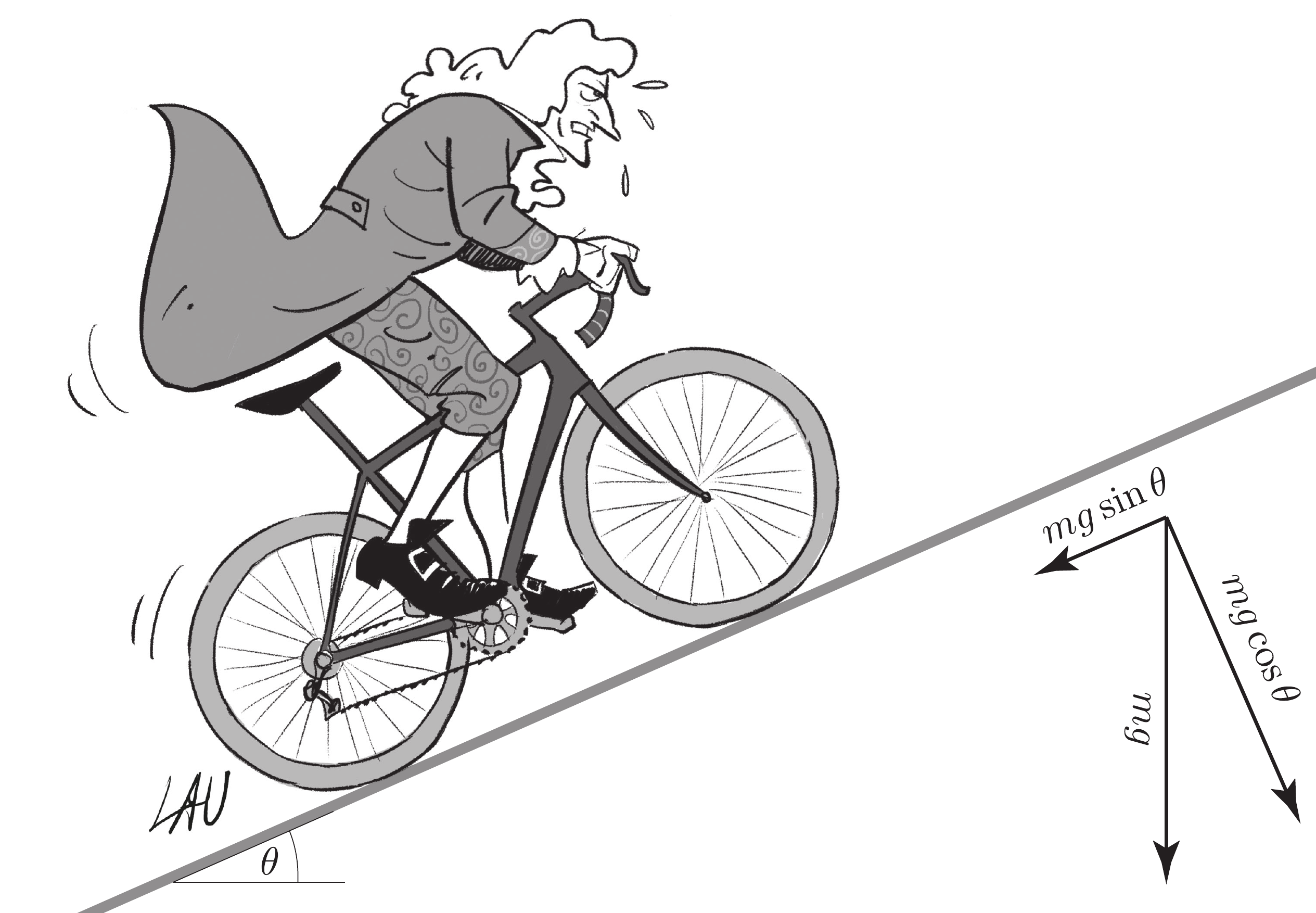}
\caption{\small Isaac Newton subject to the effects of gravity in expression~(\ref{eq:model}).}
\label{fig:FigNewton}
\end{figure}

Let us consider the following mathematical model.
\begin{align}
	\label{eq:model}
	P
	&=
	F_{\!\leftarrow}\,V_{\!\rightarrow}
	\\
	\nonumber 
	&=
	\quad\frac{
		\overbrace{\,m\,g\sin\theta\,}^\text{gravity}
		+
		\!\!\!\overbrace{\quad m\,a\quad}^\text{change of speed}
		+
		\overbrace{
			{\rm C_{rr}}\!\!\!\underbrace{\,m\,g\cos\theta}_\text{normal force}
		}^\text{rolling resistance}
		+
		\overbrace{
			\,\tfrac{1}{2}\,\eta\,{\rm C_{d}A}\,\rho\,
			(
				\!\!\underbrace{
					V_{\!\rightarrow}+w_{\leftarrow}
				}_\text{air flow speed}\!\!
			)^{2}\,
		}^\text{air resistance}
	}{
		\underbrace{\quad1-\lambda\quad}_\text{drivetrain efficiency}
	}\,V_{\!\rightarrow}\,,
\end{align}
where $F_{\!\leftarrow}$ stands for the forces opposing motion, and $V_{\!\rightarrow}$ is the ground speed of the bicycle.
$F_{\!\leftarrow}$ consists of the following quantities.
\begin{itemize}
	\item 
	$m$\,: mass of a cyclist and a bicycle%
	\footnote{\label{foot:Rotation}In this study, only translation\,---\,in contrast to rotation\,---\,of a mass is considered explicitly.
	The effects of rotation of wheels and cranks are included implicitly in coefficients~$\rm C_{d}A$\,, $\rm C_{rr}$ and $\lambda$\,.
	In Appendix~\ref{sec:RotEff}, we discuss how to accommodate explicitly the wheel rotation.}
	\item 
	$g$\,: acceleration due to gravity, whose effects are illustrated in Figure~\ref{fig:FigNewton}
	\item 
	$\theta$\,: slope
	\item 
	$a$\,: change of ground speed
	\item 
	$\rho$\,: air density,
	\begin{equation}
		\label{eq:DenAlt}
		\rho = 1.225\exp[-0.00011856\,h]\,,
	\end{equation}
	where $h$ is the altitude above the sea level%
	\footnote{In this study, we do not consider air density as a function of humidity or changeable atmospheric pressure.}
	\item 
	$w_{\leftarrow}$\,: wind component opposing the motion
	\item 
	$\rm C_{rr}$\,: unitless rolling-resistance coefficient; in a manner analogous to the friction on the plane inclined by~$\theta$, $\rm C_{rr}$ is a proportionality constant between the maximum force,~$mg$\,, and the force normal to the surface,~$mg\cos\theta$
	\item 
	$\rm C_{d}A$\,: air-resistance coefficient;%
	\footnote{\label{foot:dimensional}A formulation of $\rm C_{d}A$ is presented in Appendix~\ref{sec:AirResCoeff}.}
	a product of a unitless drag coefficient,~$\rm C_{d}$\,, and a frontal surface area, whose units are $\rm m^2$
	\item 
	$\eta$\,: a unitless quantity whose absolute value is equal to one and that ensures the proper sign for the tailwind effect,
	\begin{equation*}
\eta
={\rm sgn}(V_{\!\rightarrow}+w_\leftarrow)
=\dfrac{V_{\!\rightarrow}+w_{\leftarrow}}{\left|V_{\!\rightarrow}+w_{\leftarrow}\right|}
	\end{equation*}
	\item 
	$\lambda$\,: unitless drivetrain-resistance coefficient to account for the loss of power between the power meters and the propelling rear wheel; if power meters are in the pedals, $\lambda$ includes the resistance of bottom bracket, chain, rear sprocket and rear-wheel hub; it also includes losses due to the flexing of the frame; if power meters are in the rear-wheel hub,~${\lambda\approx 0}$
\end{itemize}
The four summands in the numerator of expression~(\ref{eq:model}) are forces to account for
\begin{itemize}
	\item
	change in elevation: increases the required power if $\theta>0$\,, decreases if $\theta<0$ and has no effect if $\theta=0$\,; it is associated with the change in potential energy,
	\item
	change in speed: increases the required power if $a>0$\,, decreases if $a<0$ and has no effect if $a=0$\,; it is associated with the change in kinetic energy, which is not lost unless the rider brakes,
	\item
	rolling resistance: increases the required power,
	\item 
	air resistance:%
	\footnote{In this study, we consider only the effect of the translation speed,~$V_{\!\rightarrow}$\,, upon the air resistance.
	 Effects due to rotation of wheels are discussed in Appendix~\ref{sec:AirResRot}.}
	increases the required power if the speed of the air flow relative to the cyclist is positive, $(V_{\!\rightarrow}+w_{\leftarrow})>0\implies\eta=1$\,, decreases if $(V_{\!\rightarrow}+w_{\leftarrow})<0\implies\eta=-1$ and has no effect if $(V_{\!\rightarrow}+w_{\leftarrow})=0$\,.
\end{itemize}
Similar models\,---\,exhibiting a satisfactory empirical adequacy\,---\,are used in other studies~\citep[e.g.,][]{MartinEtAl1998}.
\section{Extraction of resistance coefficients}
\label{sec:NumEx}
In general, obtaining a unique result for the values of the resistance coefficients by minimizing the misfit between the right-hand side of expression~(\ref{eq:formula}), which represents measurements, and the right-hand side of expression~(\ref{eq:model}), which represents a model, is impossible.
Different combinations of values give the same result.
Also, the misfit function might have several minima, with the global one not necessarily localized in the region for which the values have any physical interpretations.
For instance, as stated by \citet{Chung2012},
\begin{quote}
Remember, $\rm C_{d}A$ is an area. 
You can't have negative area.	
\end{quote}
Moreover, the optimization relies on measurements, which are subject to experimental errors, including limitations of the GPS accuracy.

To accommodate these issues in our search for the values of ${\rm C_{d}A}$\,, ${\rm C_{rr}}$ and $\lambda$\,, we group the values of the ground speed,~$V_{\!\rightarrow}$\,, illustrated in Figure~\ref{fig:FigVel}, in intervals of~$0.1$\,.
There are thirty-three such intervals, whose centres range from $8.9$ to $12.7$\,.
They contain five-hundred-and-one values of speed.
To avoid spurious results, groups are restricted to those that contain at least five values.
The mode is $9.8$\,; it is represented by twenty-five values. 
This approach stabilizes the search, by smoothing the measurements through averaging them over these intervals, and results in statistical information that gives an insight into the uncertainty of obtained results.

\begin{figure}
	\centering
	\includegraphics[scale=0.55]{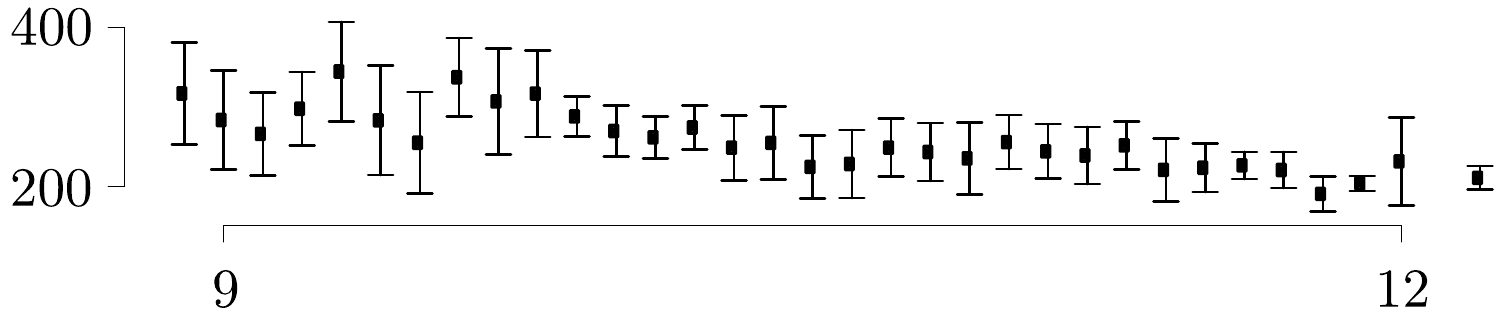}
	\caption{\small
		Power}
	\label{fig:FigPowerErr}
\end{figure}

\begin{figure}
	\centering
	\includegraphics[scale=0.55]{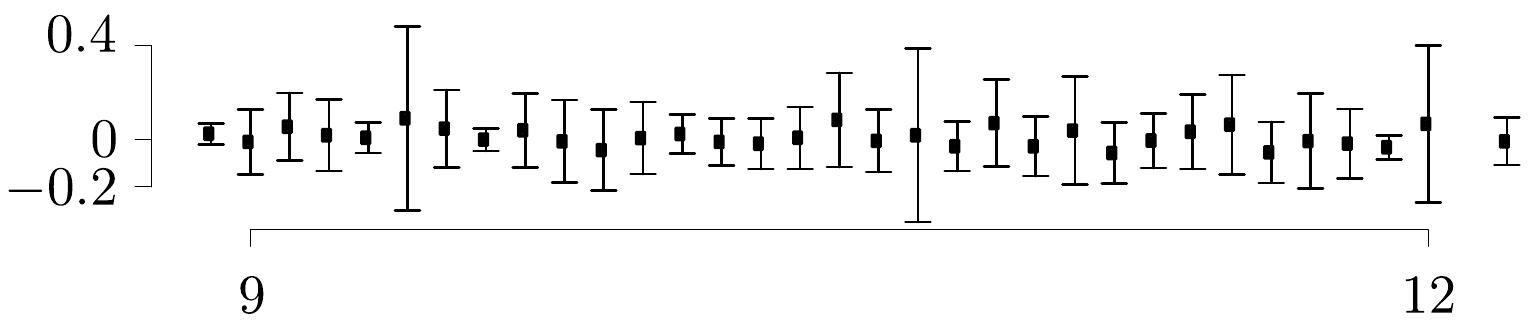}
	\caption{\small Change of speed}
	\label{fig:FigAccVel}
\end{figure}

\begin{figure}
	\centering
	\includegraphics[scale=0.55]{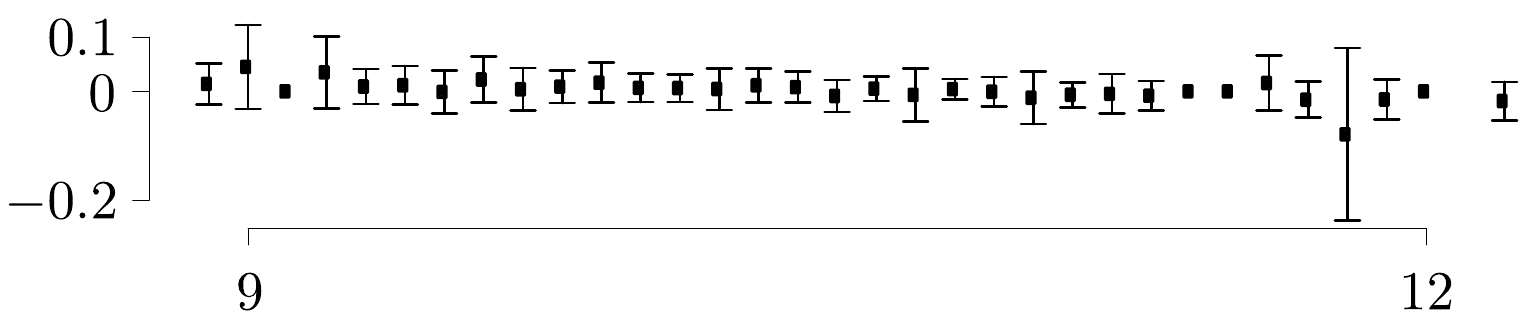}
	\caption{\small Slope}
	\label{fig:FigIncVel}
\end{figure}

\begin{figure}
	\centering
	\includegraphics[scale=0.55]{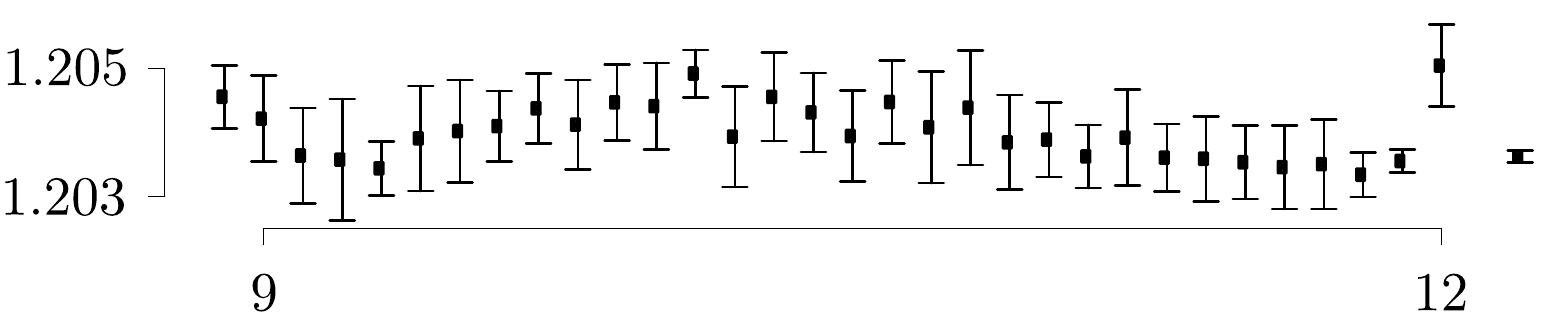}
	\caption{\small Air density}
	\label{fig:FigDenAir}
\end{figure}

The measured power\,---\,within such a grouping\,---\,is presented in Figure~\ref{fig:FigPowerErr}; therein, standard deviations are illustrated by error bars.
To provide the values required for model~(\ref{eq:model}) , we require $a$\,, $\theta$ and $\rho$\,, for each group.
They are obtained from the GPS measurements: $a$ and $\theta$ as the temporal and spatial derivatives of the measured speed and altitude, respectively, and $\rho$ by using expression~(\ref{eq:DenAlt}); these values are illustrated in Figures~\ref{fig:FigAccVel}--\ref{fig:FigDenAir}, with standard deviations illustrated by error bars.

The average change of speed\,---\,over the entire segment\,---\,is $\overline{a}=0.006922\pm0.1655$\,, which indicates a steady tempo.
The average slope\,---\,over the entire segment\,---\,is~$\overline{\theta} = 0.002575\pm0.04027$\,, which indicates a flat course.
The average air density\,---\,over the entire segment\,---\,is~$\overline\rho=1.204\pm 0.0007652$\,, which is consistent with constant atmospheric conditions.

To estimate the values of ${\rm C_{d}A}$\,, ${\rm C_{rr}}$ and $\lambda$\,, we write expression~(\ref{eq:model}) as
\begin{equation}
	\label{eq:misfit}
	f
	=
	P
	-
	\underbrace{
		\frac{
			mg\sin\theta\,V_{\!\rightarrow}
			+
			m\,a\,V_{\!\rightarrow}
			+
			{\rm C_{rr}}mg\cos\theta\,V_{\!\rightarrow}
			+
			\tfrac{1}{2}\,\eta\,{\rm C_{d}A}\,\rho
			\left(V_{\!\rightarrow}+w_{\leftarrow}\right)^{2}
			V_{\!\rightarrow}
		}{
			1-\lambda
		}
	}_{F_{\!\leftarrow}
	V_{\!\rightarrow}}
	\,,
\end{equation}
and minimize the misfit,~$\min f$\,.
The grouped values, with their standard deviations, are used as inputs for a local optimization.
Each group is treated separately and, hence, the statistics of its input parameters is different than for the entire set. 
In view of the expected values, a starting point for a local optimization is ${\rm C_{d}A}=0.3$\,, ${\rm C_{rr}}=0.005$ and $\lambda=0.035$\,.
Also, $m=111$\,, $g=9.81$ and $w_{\leftarrow}$ is set to zero.

\begin{figure}
	\centering
	\includegraphics[scale=0.7]{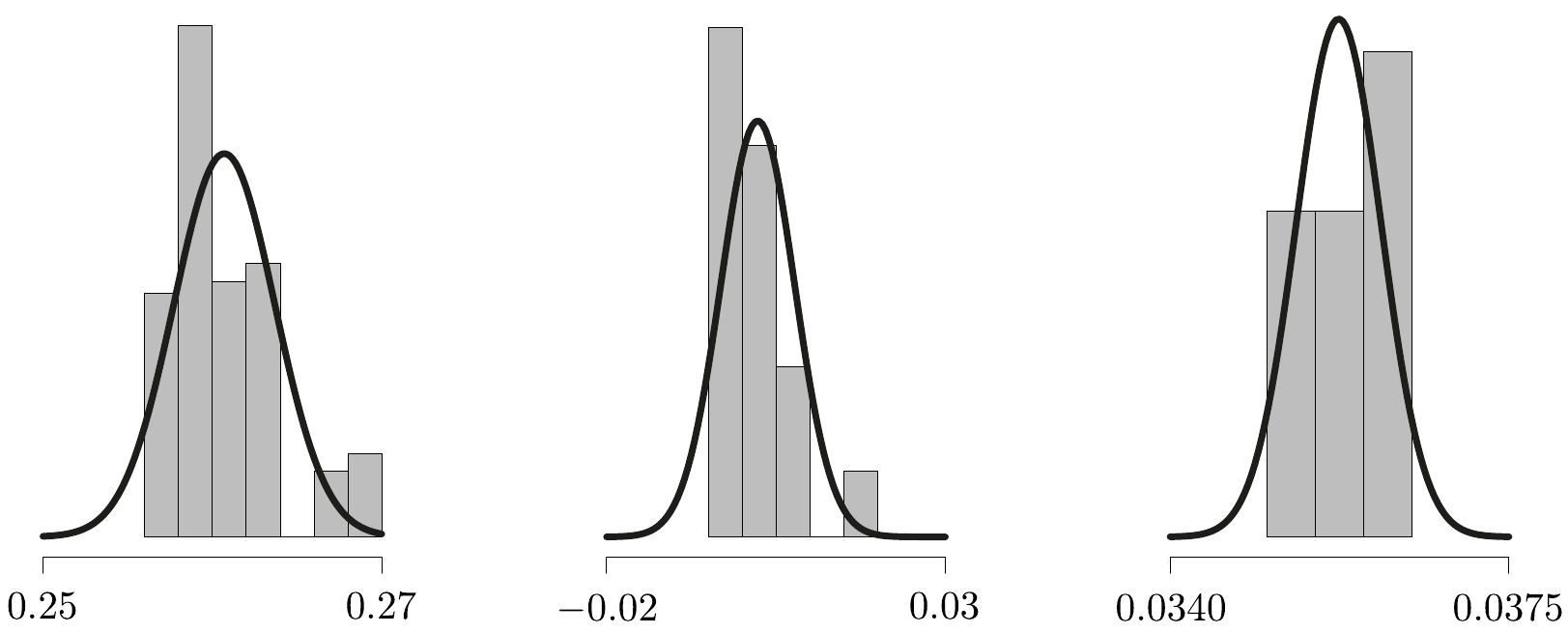}
	\caption{\small
		Optimal values; 
		left-hand plot:~${\rm C_{d}A}=0.2607\pm0.002982$\,; 
		middle plot:~${\rm C_{rr}}=0.00231\pm0.005447$\,; 
		right-hand plot:~$\lambda=0.03574\pm0.0004375$}
	\label{fig:FigParFlat}
\end{figure}

\begin{figure}
	\centering
	\includegraphics[scale=0.7]{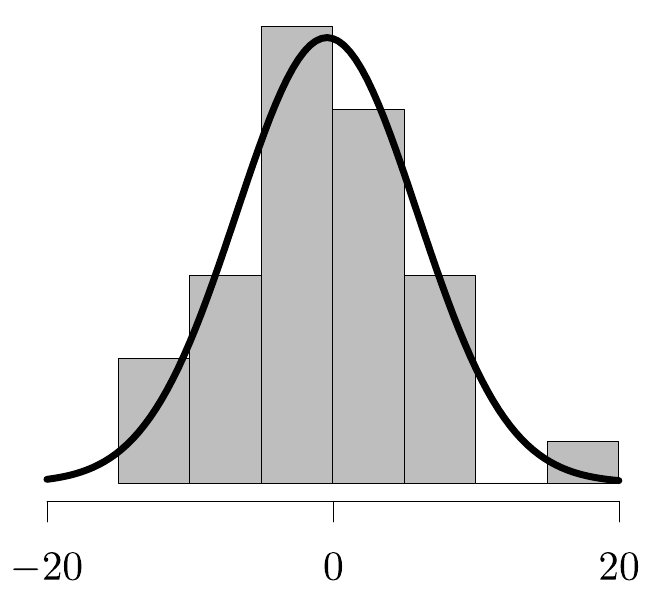}
	\caption{\small 
		Misfit of equation~(\ref{eq:misfit}): $f=0.4137\pm6.321$
	}
	\label{fig:Figfflat}
\end{figure}

The process is repeated ten thousand times.
The input values are perturbed in accordance with their Gaussian distributions, since\,---\,according to the central limit theorem\,---\,measurements affected by many independent processes tend to approximate such a distribution. 
We obtain optimal values with their standard deviations, 
\begin{equation*}
	{\rm C_{d}A} = 0.2607\pm0.002982
	\,,\,\,
	{\rm C_{rr}} = 0.00231\pm0.005447
	\,,\,\,
	\lambda = 0.03574\pm0.0004375
	\,,
\end{equation*}
shown in Figure~\ref{fig:FigParFlat}.
As illustrated in Figure~\ref{fig:Figfflat}, these values result in a satisfactory minimization of misfit for expression~(\ref{eq:misfit}).                     
Using these values, together with the average values,  over the entire segment, assuming $\overline{a}=\overline{\theta}=w_{\leftarrow}=0$\,, letting $m=111$\,, $g=9.81$\,, we obtain, in accordance with expression~(\ref{eq:model}), $P = 255.315$\,, which is consistent with $\overline{P}=258.8$\,, stated in the caption of Figure~\ref{fig:FigPower}.
\section{Discussion and conclusions}
As stated in the subtitle, the main purpose of this article is a study of estimating effects of air, rolling and drivetrain resistance.
Herein, restricting the study to a range of speeds, as illustrated in Figure~\ref{fig:FigVel}, is consistent with time trialing on a flat course.
Also, it excludes sporadic events, such as tight corners and brief mechanical failures, for which braking or stopping diminishes the quality of estimates of $\rm C_{rr}$ and $\lambda$\,, and during which an aerodynamic position is not maintained, thus diminishing the quality of estimating~$\rm C_{d}A$\,.
Furthermore, a restricted range enhances the validity of treating $\rm C_{d}A$\,, $\rm C_{rr}$ and $\lambda$ as constants, even though, in general, they are functions of speed.

For flat courses, the force of the air resistance, which is expressed by the fourth summand of expression~(\ref{eq:model}), is proportional to the square of speed and, hence, becomes dominant.
Hence, $\rm C_{d}A$ plays a significant role, in contrast to steep climbs, for which the numerator is dominated by the first summand, which accounts for the force against gravity.

The sensitivity to the force of gravity can be quantified by studying relations among $m$\,, $\theta$ and $V_{\!\rightarrow}$\,.
For instance, for any cyclist\,---\,with a given power output\,---\,the dependence of speed on mass is greater on a climb than on a flat course, and this dependence can be quantified, using $\partial V_{\!\rightarrow}/\partial m$ and the implicit function theorem, discussed in Appendix~\ref{sec:ImpFunThm}.

In estimating the effects of the air, rolling and drivetrain resistance, we recognize that the right-hand side of expression~(\ref{eq:model}), which is a forward problem, invokes $\rm C_{d}A$\,, $\rm C_{rr}$ and $\lambda$ with their independent physical meanings.
The misfit minimization of equation~(\ref{eq:misfit}), $\min f$\,, however\,---\,for which the left-hand side is the measured value and the right hand side is the retrodiction of a model\,---\,treats $\rm C_{d}A$\,, $\rm C_{rr}$ and $\lambda$ as adjustable parameters; for instance, $\rm C_{d}A$ and $\rm C_{rr}$\,, which\,---\,as physical quantities\,---\,are independent of one another, become inversely proportional to one another, for $\eta=1$\,.
In general, following the implicit function theorem, this inverse relation is
\begin{equation*}
	\dfrac{\partial\,{\rm C_{d}A}}{\partial\,{\rm C_{rr}}}
	=
	-\dfrac{
		\dfrac{\partial f}{\partial\,{\rm C_{rr}}}
	}{
		\dfrac{\partial f}{\partial\,{\rm C_{d}A}}
	}
	=
	-\dfrac{
		m\,g\cos\theta
	}{
		\tfrac{1}{2}\,\eta\,\rho
		\left(V_{\!\rightarrow}+w_{\leftarrow}\right)^{2}
	}\,,
\end{equation*}
where $f$ is given in expression~(\ref{eq:misfit}); in the present study,
\begin{equation*}
	\dfrac{\partial\,{\rm C_{d}A}}{\partial\,{\rm C_{rr}}}
	=
	-\dfrac{
		2\,m\,g
	}{
		\rho\,{V_{\!\rightarrow}}^{2}
	}
	=
	{-16.3745}
	<
	0
	\,.
\end{equation*}
Relations between the rates of change of any two quantities in expression~(\ref{eq:misfit}) can be insightful in examining the behaviour of a model.
To study the performance of a cyclist, as opposed to behaviours of a model, only a few among them are pertinent; others\,---\,such as $\partial{\rm C_{d}A}/\partial\rm C_{rr}$\,---\,are not endowed with a physical meaning.

Maintaining the physical meaning of $\rm C_{d}A$\,, $\rm C_{rr}$ and $\lambda$ remains a challenge.
To extract $\lambda$\,, one might consider a statement of \citet{Chung2012}.
\begin{quote}
	Many models include a term for overall drivetrain efficiency,~$\eta$\,,%
	\footnote{herein, $\eta\equiv 1-\lambda$}
	but all of the data files I'm looking at come from Power Taps%
	\footnote{A power meter in the rear hub}
	which, in theory, should be downstream of drivetrain losses, i.e.,~$\eta = 1$\,.
	If you have an SRM, which measures power at the crank (i.e., upstream of drivetrain losses), you will want to decide how to model drivetrain losses.
	Martin et al. presumed a fixed percentage loss of 2.3\% of power (i.e.,~$\eta = 0.977$). Other choices might include a fixed wattage loss, or loss with two components: a fixed amount and a fixed percentage.	
\end{quote}
\noindent The case of $\lambda=0$\,, used by \citet{Chung2012}, increases the stability of extracting the remaining two coefficients thanks to the disappearance of the denominator in expression~(\ref{eq:model}).
Otherwise, a division of the entire expression is a scaling that contributes to nonuniqueness.

However, even if $\lambda=0$\,, ``prying apart $\rm C_{rr}$ and $\rm C_{d}A$"\,, in words of \citet{Chung2012}, remains a challenge, for which he suggests a scenario wherein
\begin{quote}
	[i]f we test using the same tires and tubes on the same roads on the same day at the same pressure then $\rm C_{rr}$ is a constant and we can concentrate on estimating changes in $\rm C_{d}A$\,.
\end{quote}
\noindent An estimate of individual values of $\rm C_{d}A$\,, $\rm C_{rr}$ and $\lambda$\,, as physical quantities or, at least, their relative changes is an important aspect of our work.
We wish to ensure a sufficient accuracy to examine effects\,---\,on $\rm C_{d}A$\,---\,of aerodynamic equipment and riding position, as well as the efficiency of drafting for a team time trial.
Also, we wish to examine the effects\,---\,on $\rm C_{rr}$\,---\,of tire width and pressure.
Furthermore, we wish to examine the effects\,---\,on $\lambda$\,---\,of different sizes of the chainring and cog that result in the same ratio, as well as a the efficiency of a fixed gear versus freewheel.
\section*{Acknowledgements}
We wish to acknowledge Len Bos and Yves Rogister, for our fruitful discussions, David Dalton, for his scientific editing and proofreading, Elena Patarini, for her graphic support, and Roberto Lauciello, for his artistic contribution.
Furthermore, we wish to acknowledge Favero Electronics, including their Project Manager, Francesco Sirio Basilico, and R\&D engineer, Renzo Pozzobon, for inspiring this study by their technological advances and for supporting this work by insightful discussions and providing us with their latest model of Assioma Duo power meters.
\bibliographystyle{spbasic}
\bibliography{DSSbici1_arXiv.bib}
\begin{appendix}
\section{On effects of averaging pedal speed per revolution for power calculations}
\label{sec:AppendixA}
\setcounter{equation}{0}
\setcounter{figure}{0}
\renewcommand{\theequation}{\Alph{section}.\arabic{equation}}
\renewcommand{\thefigure}{\Alph{section}\arabic{figure}}
\subsection{Preliminary remarks}
Since, in expression~(\ref{eq:formula}), $v_{\circlearrowright}$ is proportional to the cadence, it is common to simplify circumferential speed measurements by considering only the cadence.
This means that---instead of measuring the speed instantaneously along the revolution---the measurement is performed only once per revolution and the resulting average is used in subsequent calculations.
In this appendix, we examine the effects of such a simplification.

Referring to an expression equivalent to expression~(\ref{eq:formula})---that invokes torque and angular velocity instead of the force and circumferential speed---\citet{Favero2018} state that
\begin{quote}
	[t]he torque/force value is usually measured many times during each rotation, while the angular velocity variation is commonly neglected, considering only its average value for each revolution.
	$[\,\ldots\,]$
	Favero Electronics, to ensure the maximum accuracy of its power meters in all pedaling conditions, decided to research to what extent the variation of angular velocity during a rotation affects the power calculation.	
\end{quote}
To examine the effect of including speed variation during a revolution, let us consider the following formulation to gain  analytical insights into the empirical results obtained by~\citet{Favero2018}.
\subsection{Formulation}
\begin{figure}
	\centering
	\includegraphics[scale=0.5]{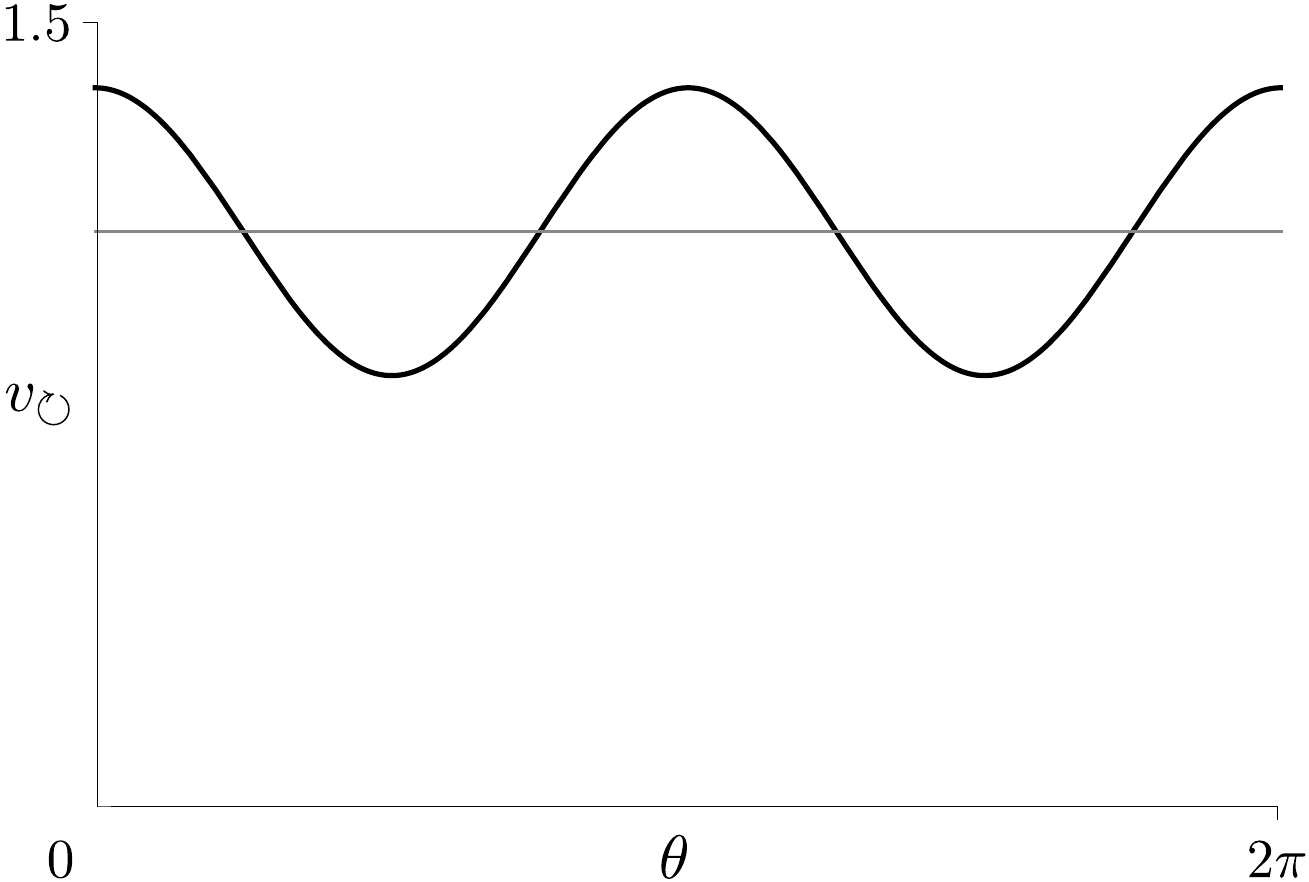}
	\caption{\small
		Circumferential speeds corresponding to expression~(\ref{eq:speed}): $a=0.25$\,, $r=0.175$ and $a=0$\,, $r=0.175$\,;  the former shown in black and the latter in grey
	}
	\label{fig:FigSpeed}
\end{figure}

Consider a pedal whose revolution takes one second; hence, its circumferential speed is
\begin{equation}
	\label{eq:speed}
	v_{\circlearrowright}(\theta)
	=
	v_{0}\left(1+a\cos(2\theta)\right)
	\,,\qquad
	\theta\in(0,2\pi]
	\,,
\end{equation}
where $v_{0}=2\pi r/1$\,, $r$ is the crank length and $\theta$ is the angle.
Expression~(\ref{eq:speed}) is illustrated in Figure~\ref{fig:FigSpeed}.

\begin{figure}
	\centering
	\includegraphics[scale=0.5]{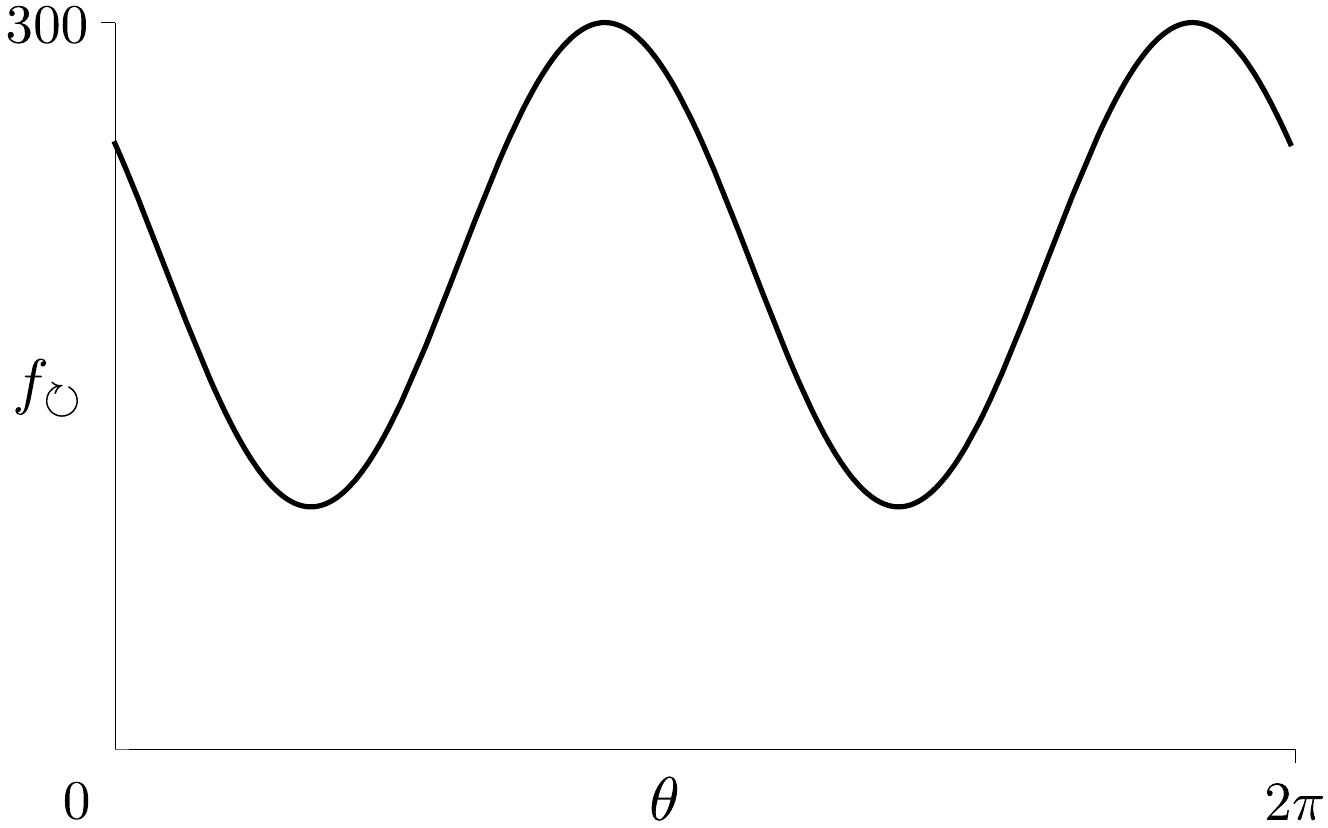}
	\caption{\small
		Applied force corresponding to expression~(\ref{eq:force}): $b=0.5$\,, $c=\pi/6$\,, $f_{0}=200$
	}
	\label{fig:FigForce}
\end{figure}

Assume the magnitude of the tangential component of force applied to both pedals during this revolution to be
\begin{equation}
	\label{eq:force}
	f_{\circlearrowright}(\theta)
	=
	f_{0}\left(1+b\cos(2(\theta+c))\right)
	\,,\qquad
	\theta\in(0,2\pi]\,,	
\end{equation}
where $f_{0}$ is a constant and $c$ is an angular shift between $v_{\circlearrowright}$ and $f_{\circlearrowright}$\,, which is a constant whose units are radians.
Expression~(\ref{eq:force}) is illustrated in Figure~\ref{fig:FigForce}.

In accordance with expression~(\ref{eq:formula}), the instantaneous power, at $\theta$\,, is $P_{\circlearrowright}(\theta)=f_{\circlearrowright}(\theta)\,v_{\circlearrowright}(\theta)$\,, and the average power over the revolution is
\begin{equation}
	\label{eq:InsPow}
	\overline{P}_{\circlearrowright}
	=
	\dfrac{1}{2\pi}
	\int\limits_{0}^{2\pi}
	P_{\circlearrowright}(\theta)
	{\,\rm d}\theta
	=
	\left(2+a\,b\cos(2\,c)\right)\pi r f_{0}\,.
\end{equation}
If we consider the average value of speed,
\begin{equation}
	\label{eq:velocity}
	\overline{v}_{\circlearrowright}
	=
	\dfrac{1}{2\pi}
	\int\limits_{0}^{2\pi}
	v_{\circlearrowright}(\theta)
	{\,\rm d}\theta
	=
	2\pi r
	\,,	
\end{equation}
then,
\begin{equation}
	\label{eq:AvePow}
	\overline{P}_{\circlearrowright}
	=
	\dfrac{1}{2\pi}
	\int\limits_{0}^{2\pi}
	f_{\circlearrowright}(\theta)\,\overline{v}_{\circlearrowright}
	{\,\rm d}\theta
	=
	2\pi r f_{0}
	\,,	
\end{equation}
over one revolution;  $a$\,, $b$ and $c$ have no effect on $\overline{P}_{\circlearrowright}$\,.
Examining expressions~(\ref{eq:InsPow}) and (\ref{eq:AvePow}), we see that the former reduces to the latter if $a\,b=0$ or if $c=\pi/4$ or $c=3\pi/4$\,.
Otherwise, the power over a revolution---based on the instantaneous speed---is different from the power based on the speed averaged for each revolution.
One might note that expression~(\ref{eq:AvePow}) can be also obtained as the product of expression~(\ref{eq:velocity}) and
\begin{equation}
	\label{eq:BarF}
	\overline{f}_{\circlearrowright}
	=
	\frac{1}{2\pi}
	\int\limits_{0}^{2\pi}
	f_{0}\left(1+b\cos(2(\theta+c)\right)
	{\rm d}\theta
	=
	f_{0}
	\,,
\end{equation}
which is the average force per revolution that results from expression~(\ref{eq:force})\,.	
\begin{figure}
	\centering
	\includegraphics[scale=0.5]{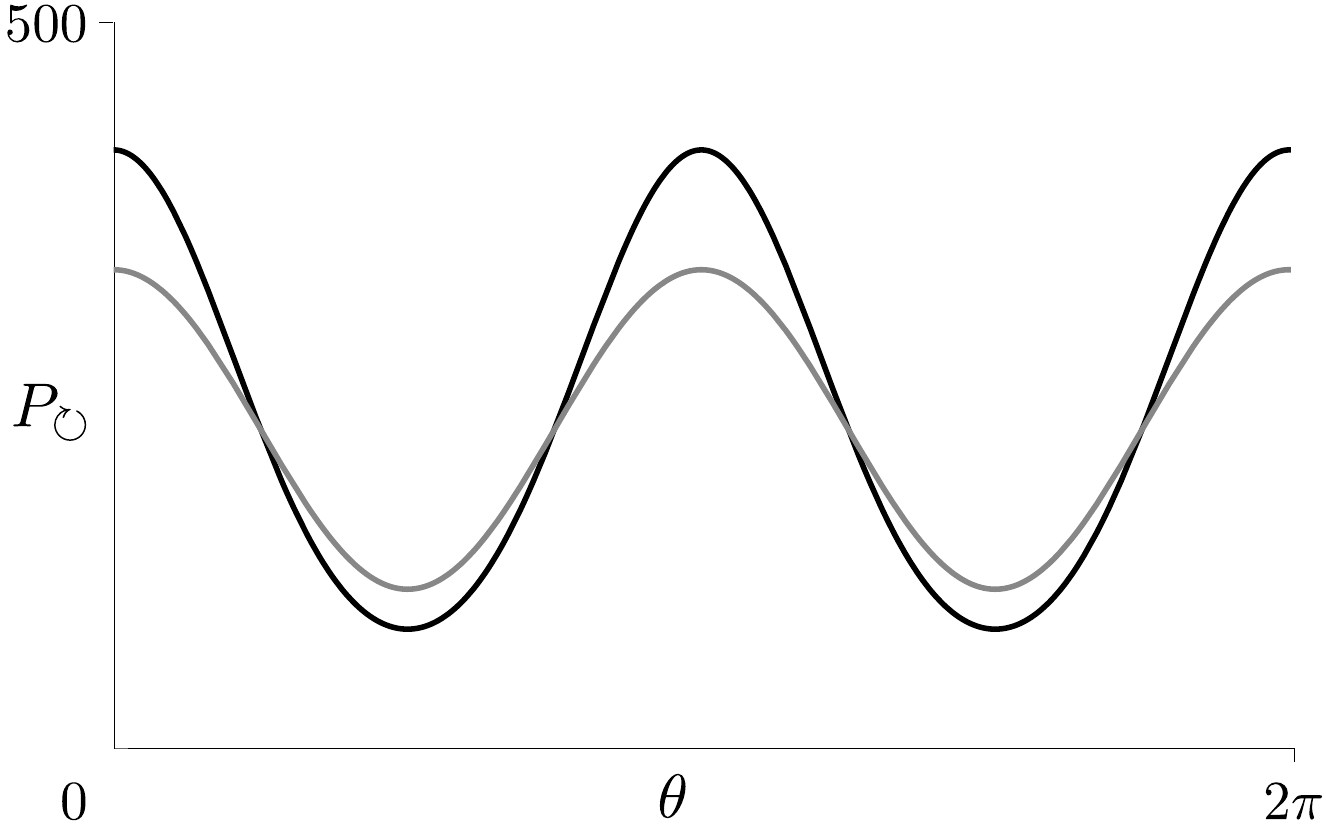}
	\caption{\small
		Instantaneous power corresponding to expressions~(\ref{eq:InsPow}) and (\ref{eq:AvePow}); the former shown in black and the latter in grey, with averages of $227$ and $220$\,, respectively
	}
	\label{fig:FigPowerRot}
\end{figure}

The integrands of expressions~(\ref{eq:InsPow}), with $c=0$\,, and (\ref{eq:AvePow}) are illustrated in Figure~\ref{fig:FigPowerRot}.
Therein, following expressions~(\ref{eq:speed}) and (\ref{eq:force}), the integrand of expression~(\ref{eq:InsPow})  is
\begin{equation*}
	P_{\circlearrowright}(\theta)
	=
	f_{\circlearrowright}(\theta)\,
	v_{\circlearrowright}(\theta)
	=
	f_{0}\,(1+b\cos(2(\theta+c)))\,v_{0}(1+a\cos(2\theta))
	\,.
\end{equation*}
Invoking trigonometric identities and rearranging, we write it as
\begin{equation*}
	P_{\circlearrowright}(\theta)
	=
	f_{0}\,v_{0}
	\left(
		\underbrace{1+\dfrac{ab}{2}\cos(2c)}_{\rm constant}
		+
		\underbrace{
			\dfrac{}{}a\cos(2\theta) + b\cos(2(\theta+c))
		}_{\rm double\,frequency}
		+
		\underbrace{
			\dfrac{ab}{2}\cos(4\theta+2c)
		}_{\rm quadruple\,frequency}\,
	\right)
	\,.
\end{equation*}
However, the effect of the third summand is small enough not to appear in Figure~\ref{fig:FigPowerRot}.
For instance, if we let $c=0$\,, the double-frequency term becomes $(a+b)\cos(2\theta)$ and the quadruple frequency term becomes $\tfrac{ab}{2}\cos(4\theta)$\,.
If $a<1$ and $b<1$\,, the amplitude of the third summand is much smaller, and the appearance of the plot is dominated by the double-frequency term.
\subsection{Numerical examples}
\begin{figure}
	\centering
	\includegraphics[scale=0.5]{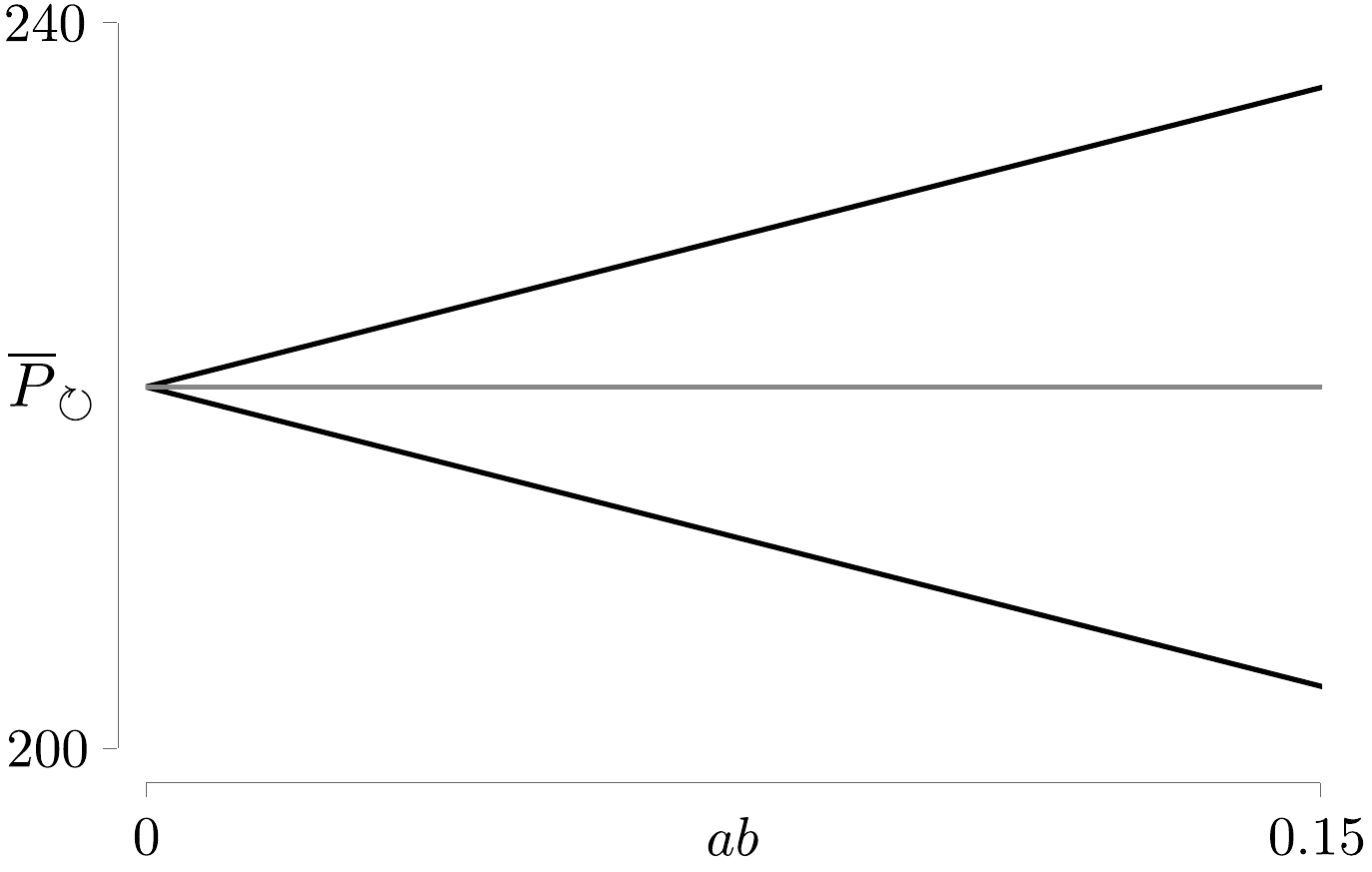}
	\caption{\small
		Average power corresponding to expressions~(\ref{eq:InsPow}) and (\ref{eq:AvePow}); the former shown in black and the latter in grey; the former depends on $ab$\,, the latter does not; the former depends on $c$\,, the latter does not; $c=0$\,, for the increasing line, $c=\pi/2$ or $c=3\pi/2$\,, for the decreasing line
	}
	\label{fig:FigAverage}
\end{figure}

If we let $a= 0.25$\,, $b = 0.5$\,, $c = 0$\,, $f_{0} = 200$\,, $v_{0}=2\pi r$ and $r=0.175$\,, expression~(\ref{eq:InsPow}) results in $234$~watts, as the average power per revolution, and expression~(\ref{eq:AvePow}) in $220$~watts.
The approach that neglects speed variations during the revolution can also overestimate the average power.
If we let $c=\pi/2$\,, expression~(\ref{eq:InsPow}) results in $\overline{P}=206$ and expression~(\ref{eq:AvePow}) remains unchanged.
These results are illustrated in Figure~\ref{fig:FigAverage}.

Expressions~(\ref{eq:speed})--(\ref{eq:BarF}) refer to a single revolution.
Hence, the values resulting from expressions~(\ref{eq:InsPow}) and (\ref{eq:AvePow}) remain the same, regardless of cadence; they are averages over one rotation.

If the pedaling is smoother, as one might expect for higher cadences, the values of $a$ and $b$ become smaller.
Since these values are smaller than unity and appear as a product, expression~(\ref{eq:InsPow}) might approach expression~(\ref{eq:AvePow}).
If we let $a=0.1$\,, $b=0.3$\,, $c=0$\,, $f_{0}=200$ and $r=0.175$\,, expression~(\ref{eq:InsPow}) results in $\overline{P}=223$\,; the value of expression~(\ref{eq:AvePow}) remains unchanged.

Furthermore, for a single revolution, there is a unique pair of force and speed that results in a power given by expressions~(\ref{eq:InsPow}) and (\ref{eq:AvePow}).
However---for a given time interval and various cadences---there are many pairs of force and speed that result in the same value of power.

\begin{figure}
	\centering
	\includegraphics[scale=0.5]{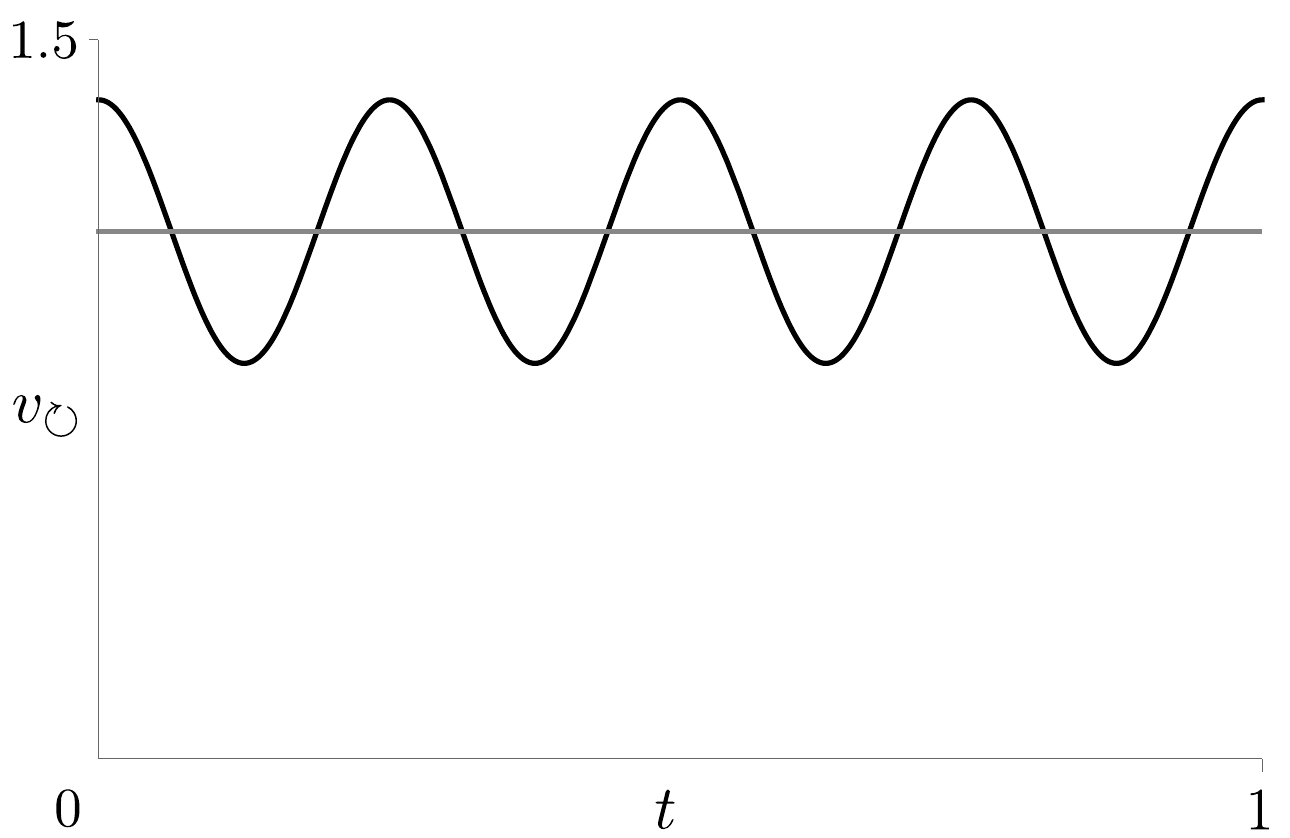}
	\caption{\small
		Circumferential speeds corresponding to expression~(\ref{eq:speed2}): $a=0.25$\,, $r=0.175$ and $a=0$\,, $r=0.175$\,;  the former shown in black and the latter in grey
	}
	\label{fig:FigRenzoOne}
\end{figure}
\begin{figure}
	\centering
	\includegraphics[scale=0.5]{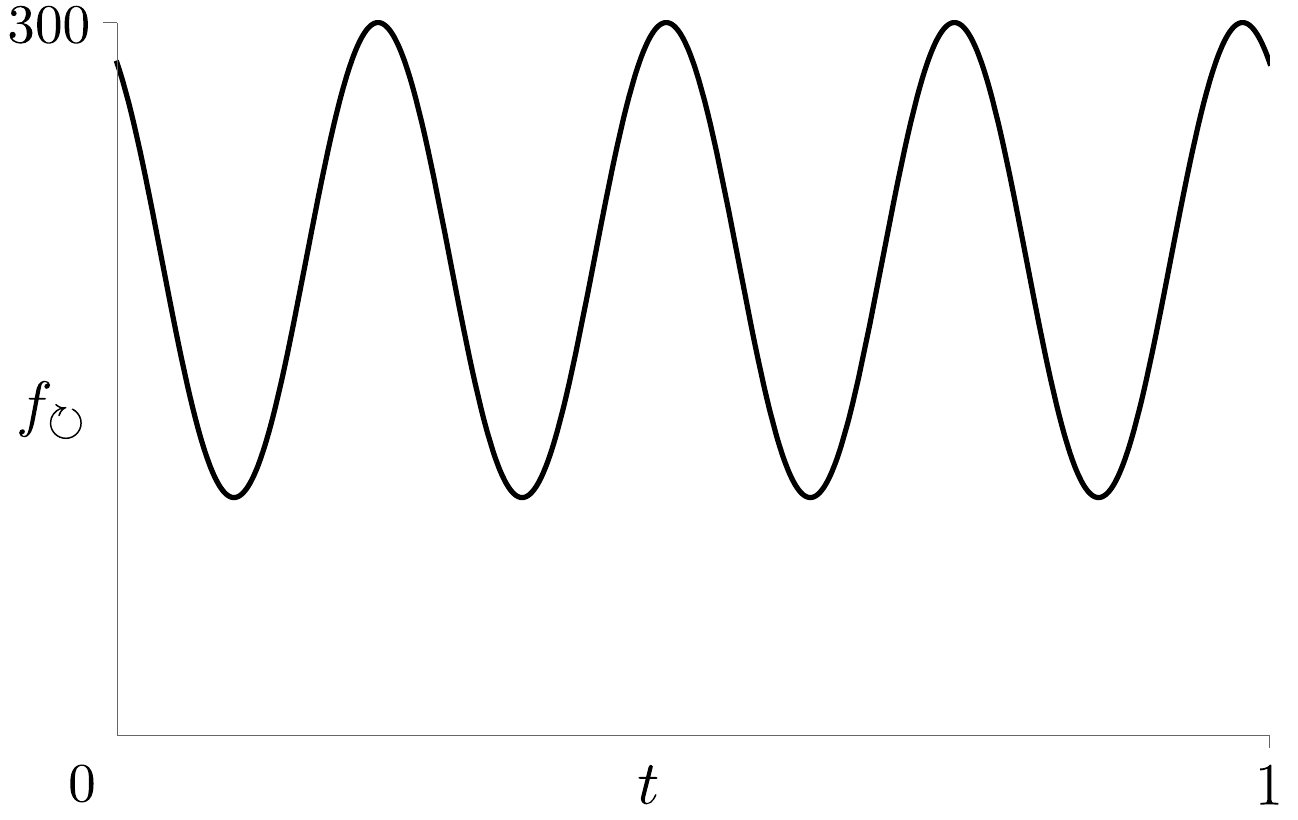}
	\caption{\small
		Applied force corresponding to expression~(\ref{eq:force}): $b=0.5$\,, $c=\pi/6$\,, $f_{0}=200$
	}
	\label{fig:FigRenzoTwo}
\end{figure}
For expression~(\ref{eq:speed}) to correspond to two revolutions per second, we modify it to be
\begin{equation}
	\label{eq:speed2}
	v_{\circlearrowright}(t)
	=
	4\pi r(1+a\cos(8\pi t))
	\,,\qquad
	t\in(0,1]
	\,,
\end{equation}
where $t$ stands for time; expression~(\ref{eq:speed2}) is illustrated in Figure~\ref{fig:FigRenzoOne}.
Accordingly, we modify expression~(\ref{eq:force}) to be
\begin{equation}
	\label{eq:force2}
	f_{\circlearrowright}(t)
	=
	f_{0}\,(1+b\cos(8\pi(t+c)))
	\,,\qquad
	t\in(0,1]
	\,,
\end{equation}
where $c$ is a time shift between $v_{\circlearrowright}$ and $f_{\circlearrowright}$\,, which is a constant whose units are seconds; expression~(\ref{eq:force2}) is illustrated in Figure~\ref{fig:FigRenzoTwo}.
Hence, expression~(\ref{eq:InsPow}) becomes
\begin{align}
	\nonumber
	\overline{P}_{\circlearrowright}
	&=
	\int\limits_{0}^{1}
	v_{\circlearrowright}(t)\,f_{\circlearrowright}(t)
	{\,\rm d}t
	=
	\int\limits_{0}^{1}
	\overbrace{
		4\pi r(1+a\cos(8\pi t))
	}^{v_{\circlearrowright}(t)}
	\overbrace{
		f_{0}(1+b\cos(8\pi(t+c)))
	}^{f_{\circlearrowright}(t)}
	{\,\rm d}t
	\\
	\label{eq:AveTime}
	&=
	2(2+ab\cos(8\pi c))\pi r f_{0}\,.
\end{align}
Examining expressions~(\ref{eq:InsPow}) and (\ref{eq:AveTime}), we see that to keep the same average power per second\,---\,with $ab=0$ or $c=0$\,---\,we need to halve the value of~$f_{0}$\,.
Otherwise, the ratio is
\begin{equation*}
	\dfrac{2+ab\cos(2\pi c)}{2\left(2 + ab\cos(8\pi c)\right)}
	\,.
\end{equation*}

In accordance with expression~(\ref{eq:InsPow}), which is tantamount to the power averaged over one second\,---\,if the cadence is one revolution a second\,---\,and given $a=0.25$\,, $b=0.5$\,, $c=0$\,, $f_{0}=200$\,, $r=0.175$\,, we have $\overline{P}_{\circlearrowright}=234$\,.
With a cadence of two revolutions a second, in accordance with expression~(\ref{eq:AveTime}), the same average power is obtained with~$f_{0}=100$\,.
Thus, among many possible pairs that result in $\overline{P}_{\circlearrowright}=234$\,, we have $(60$~rpm, $f_{0}=200)$ and $(120$~rpm, $f_{0}=100)$\,.

If $a=0$\,, in accordance with expression~(\ref{eq:InsPow}) and (\ref{eq:AvePow}), $\overline{P}_{\circlearrowright}=2\pi r f_{0}$\,, per second, and, in accordance with expression~(\ref{eq:AveTime}), $\overline{P}_{\circlearrowright}=4\pi r f_{0}$\,, per second.
Thus, to keep the same average power, we again halve the value of~$f_{0}$\,.
If the original value, at $60$~rpm, is $\overline{P}_{\circlearrowright}=234$\,, the corresponding value is calculated to be $f_{0}=213$\,, instead of~$200$\,.
This results in a different\,---\,and less accurate\,---\,pair, due to neglecting speed variation during a revolution.
\subsection{Closing remarks}
As illustrated in this appendix, there is a discrepancy between the power-meter calculations resulting from the use of the instantaneous-speed and average-speed information.
Removing this discrepancy is crucial for a variety of information that rely on power measurements, as is the case of this paper.

Let us conclude by addressing the issue of sampling with regards to the discrepancy in power calculation resulting from $\overline{v}_{\circlearrowright}$ as opposed to $v_{\circlearrowright}(\theta)$\,. 
Let us consider the gear of $54\times17$\,.
For a road bicycle, one revolution results in a development of $6.67$~metres.
Hence, for the speed of $48.1$~kilometres per hour, a full rotation corresponds to half a second, which is a high time-trial cadence of $120$~revolutions per minute.
The sampling of twice-a-second, however, is insufficient for accurate information about power.
For that reason, the Favero power meters provide the cycling computer with data that already contains information based on the instantaneous---as opposed to the average---speed.
\section{Rotation effects: moment of inertia}
\label{sec:RotEff}
\setcounter{equation}{0}
\setcounter{figure}{0}
\renewcommand{\theequation}{\Alph{section}.\arabic{equation}}
\renewcommand{\thefigure}{\Alph{section}\arabic{figure}}
To include the effect of rotation upon change of speed in the model stated in expression~(\ref{eq:model}), we consider the moment of inertia, which is $mr^2$\,, for a thin circular loop, and $mr^2/2$\,, for a solid disk, where $r$ stands for their radius.
	Relating the angular change in speed to the circumferential one\,---\,by a temporal derivative of $v=\omega r$\,, where $v$ is the circumferential speed and $\omega$ is the angular speed\,---\,the magnitudes of the corresponding rotational force are $Fr=mra$ and  $Fr=mra/2$\,, respectively; hence, $F=ma$ and $F=ma/2$ are the corresponding linear forces.
	
\begin{figure}
\centering
\includegraphics[scale=0.7]{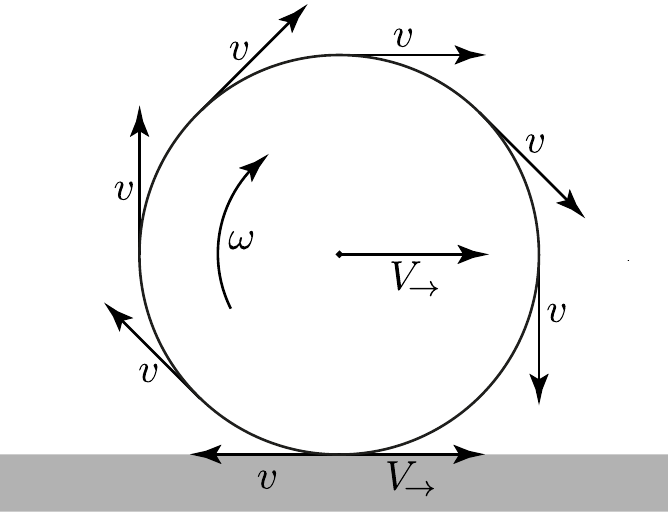} 
\caption{\small Rolling without slipping: angular speed,~$\omega$\,, circumferential speed,~$v=\omega\,r$\,, where $r$ is radius, and bicycle speed,~$V_{\!\rightarrow}=v$}
\label{fig:FigNoSlip}
\end{figure}

	In the preceding paragraph, $v=\omega r$ is the circumferential speed.
	To show that it is equal to the ground speed of the bicycle,~$V_{\!\rightarrow}$\,, let us consider the point of contact of the wheel and the road.
	The ground speed of that point is the sum of the circumferential speed of the wheel, at that point, and the speed of the bicycle.
	Since\,---\,under assumption of no slipping\,---\,the ground speed of that point is zero and the other two speeds refer to velocities in the opposite directions, we have $v=V_{\!\rightarrow}$\,.
	Thus, as illustrated in Figure~\ref{fig:FigNoSlip}, the circumferential speed of the wheel is the same as the ground speed of the bicycle; the same is true for the change of speed, as required.
	
To consider a bicycle with one standard wheel and one disk wheel, we denote the mass of the former by~$m_{\rm w}$\,, and the mass of the latter by~$m_{\rm d}$\,.
Thus, the second summand in the numerator of expression~(\ref{eq:model}), which is a linear force, becomes $(m+m_{\rm w}+m_{\rm d}/2)\,a$\,.
	
Note that $m$ contains both $m_{\rm w}$ and $m_{\rm d}$ to account for the translational and rotational effects; the former depends on the total mass and the latter on the mass of the wheels only.
For a standard wheel, $m_{\rm w}\approx 0.8$\,, and for a disk wheel,~$m_{\rm d}\approx 1.3$\,; for both~$r=0.31$\,.
\section{Air-resistance coefficient}
\label{sec:AirResCoeff}
\setcounter{equation}{0}
\setcounter{figure}{0}
\renewcommand{\theequation}{\Alph{section}.\arabic{equation}}
\renewcommand{\thefigure}{\Alph{section}\arabic{figure}}
To formulate the air-resistance force in expression~(\ref{eq:model}), we assume that it is proportional to the frontal area,~$A$\,, and to the pressure,~$p$\,, exerted by air on this area,~$F_a\propto pA$\,, where $p=\tfrac{1}{2}\rho V^2$ has a form of kinetic energy and $V=V_{\!\rightarrow}+w_\leftarrow$ is the relative speed of a cyclist with respect to the air; $p$ is the energy density per unit volume.
We can write this proportionality as
\begin{equation*}
F_a=\tfrac{1}{2}{\rm C_d}A\rho V^2\,,
\end{equation*}
where $\rm C_d$ is a proportionality constant, which is referred to as the drag coefficient.

A more involved justification for the form of the air-resistance force in expression~(\ref{eq:model}) is based on dimensional analysis \citep[e.g.,][Chapter~3]{Birkhoff}.
We consider the air-resistance force, which is a dependent variable, as an argument of a function, together with the independent variables, to write
\begin{equation*}
f(F_a,V,\rho,A,\nu)=0\,; 	
\end{equation*}
herein, $\nu$ is the viscosity coefficient.
Since this function is zero in any system of units, it is possible to express it in terms of dimensionless groups, only.

According to the Buckingham theorem \citep[e.g.,][Chapter~3, Section~4]{Birkhoff}\,---\,since there are five variables and three physical dimensions, namely, mass, time and length---\,we can express the arguments of~$f$ in terms of two dimensionless groups.
There are many possibilities of such groups, all of which lead to equivalent results.
A common choice for the two groups is
\begin{equation*}
\frac{F_a}{\tfrac{1}{2}\,\rho\,A\,V^2}\,,
\end{equation*}
which is referred to as the drag coefficient, and
\begin{equation*}
\frac{V\,\sqrt{A}}{\nu }\,,
\end{equation*}
which is referred to as the Reynolds number.
Thus, treating physical dimensions as algebraic objects, we can reduce a function of five variables into a function of two variables,
\begin{equation*}
g\left(\frac{F_a}{\tfrac{1}{2}\,\rho\,A\,V^2}\,,\,\frac{V\,\sqrt{A}}{\nu }\right)\,=\,0\,,
\end{equation*}
which we write as
\begin{equation*}
\frac{F_a}{\tfrac{1}{2}\,\rho\,A\,V^2}=h\left(\frac{V\,\sqrt{A}}{\nu }\right)\,,
\end{equation*}
where the only unknown is~$F_a$\,, and where $h$ is a function of the Reynolds number.
Denoting the right-hand side by $\rm C_d$\,, we write
\begin{equation*}
F_a=\tfrac{1}{2}{\rm C_d}\,A\,\rho\,V^2\,,
\end{equation*}
as expected.
In view of this derivation, $\rm C_d$ is not a constant; it is a function of the Reynolds number.
In our study, however\,---\,within a limited range of speeds\,---\,$\rm C_d$ is treated as a constant.
Furthermore, since $A$ is difficult to estimate, we include it within this constant, and consider $\rm C_dA$\,.
\section{Rotation effects: air resistance}
\label{sec:AirResRot}
\setcounter{equation}{0}
\setcounter{figure}{0}
\renewcommand{\theequation}{\Alph{section}.\arabic{equation}}
\renewcommand{\thefigure}{\Alph{section}\arabic{figure}}
To include the effect of air resistance of rotating wheels in the model stated in expression~(\ref{eq:model}), another summand is to be introduced to the numerator, namely,
\begin{equation*}
\tfrac{1}{2}{\rm C_w}\pi r^2\rho\,(V_{\!\rightarrow}+w_{\leftarrow})^2\,,	
\end{equation*}
where $r$ is the wheel radius.
Such a summand is formulated by invoking dimensional analysis in a manner analogous to the one presented in Appendix~\ref{sec:AirResCoeff}.

To combine rotational air resistance with the translational one, we use $v=\omega r$\,, where $v$ is the circumferential speed and $\omega$ is the angular speed, and the fact that\,---\,as discussed in Appendix~\ref{sec:RotEff} and illustrated in Figure~\ref{fig:FigNoSlip}\,---\,the circumferential speed, under the assumption of rolling without slipping, is the same as the ground speed of the bicycle,~$V_{\!\rightarrow}$\,.

Considering two standard wheels, we write
	\begin{equation}
		\label{eq:model2}
		P
		=
		\frac{
			mg\sin\theta 
			+
			(m+2\,m_{\rm w})\,a 
			+
			{\rm C_{rr}}m g\cos\theta 
			+
			\tfrac{1}{2}\eta\,\rho
			\,(
				2\,{\rm C_w}\overbrace{\pi r^2}^{A_\circ}
				+
				{\rm C_{d}A_f}
			)
			\left(V_{\!\rightarrow} + w_{\leftarrow}\right)^2
		}{
			1-\lambda
		}
		V_{\!\rightarrow}
		\,;	
	\end{equation}
herein, in contrast to expression~(\ref{eq:model}) and as discussed in Appendix~\ref{sec:RotEff}, the change of speed, expressed by the second summand, contains effects of the moment of inertia due to rolling wheels.
The air resistance, expressed by the fourth summand, distinguishes between the air resistance due to translation of a bicycle from the air resistance due to its rolling wheels.
$\rm A_f$ is the entire frontal area and $\rm A_\circ$ is the wheel side area.%
\footnote{Using expression~(\ref{eq:model2}) and the implicit function theorem, ${\partial\,{\rm C_dA_f}/\partial\,{\rm C_wA_\circ}=-2}$\,, which is indicative of the behaviour of a model; there is no physical relation between $\rm C_dA_f$ and $\rm C_wA_\circ$\,.
We expect, $\rm C_wA_o \ll C_dA_f$\,; however, the optimization programs treat them as two adjustable parameters of equal importance.}
An examination of the effect of two different wheels requires the introduction of two coefficients, one for each wheel.

In this study, the quality of available information renders the extraction of values for the resistance coefficients difficult.
Even though the data obtained from the power meter has high accuracy, the data based on the GPS measurements introduces the uncertainty that renders an accurate extraction of even three parameters a numerical challenge.
Extraction of four or five parameters requires further studies and, above all, more reliable data.

In the meantime, we can consider forward estimates, such as gaining an insight into the effect of a disk wheel.
Following expression~(\ref{eq:model2})\,---\,under windless conditions,~$w_{\leftarrow}=0$\,, on a flat course,~$\theta=0$\,, and with a steady tempo,~$a=0$\,---\,we write the required powers as
\begin{equation*}
	P_{\rm n}=\dfrac{{\rm C_{rr}}mg+\tfrac{1}{2}\,\rho\left(2\,{\rm C_{w_n}A_w}+{\rm C_{d}A_f}\,\right)V_{\!\rightarrow}{}^2}{1-\lambda}V_{\!\rightarrow}	
	\end{equation*}
and
\begin{equation*}
	P_{\rm d}=\dfrac{{\rm C_{rr}}mg+\tfrac{1}{2}\,\rho\left(({\rm C_{w_n}}+{\rm C_{w_d}}){\rm A_w}+{\rm C_{d}A_f}\,\right)V_{\!\rightarrow}{}^2}{1-\lambda}V_{\!\rightarrow}\,,	
	\end{equation*}
where we distinguish between the drag coefficients of a normal wheel and a disk wheel.
The difference in required power is
\begin{equation*}
	\Delta P
	=
	\dfrac{{\rm C_{w_n}}-{\rm C_{w_d}}}{2\left(1-\lambda\right)}\,{\rm A_w}\,\rho\,{V_{\!\rightarrow}}^{3}\,.
\end{equation*}
Letting ${\rm C_{w_n}}\approx 0.05$ and ${\rm C_{w_d}}\approx 0.035$~\citep{GreenwellEtAl1995} means that, for a standard wheel, ${\rm C_wA_\circ}\approx 0.015$\,, and for a disk wheel,~${\rm C_wA_\circ}\approx 0.01$\,.
Both values are significantly smaller than $\rm C_dA_f$\,, as expected.
Letting $r=0.31$\,, $\rho=1.204$\,, $\lambda=0.03574$\,, we obtain $\Delta P\approx0.0028V_{\!\rightarrow}{}^3$\,.
For $\overline V_{\!\rightarrow}=10.51$\,, we have $\Delta P\approx 3.3$\,.
Thus, for the present study, in the neighbourhood of $\overline P=258.8$\,, the replacement of a regular wheel by a disk wheel results in the decrease of required power of $\Delta P/\overline P\approx 1.3\%$\,, to maintain the same speed.
The disadvantages of the weight of a disk wheel almost disappear for a flat course,~$\theta=0$\,, and a steady tempo,~$a=0$\,; $m$ remains in the third summand, only.
\section{Rates of change}
\label{sec:ImpFunThm}
\setcounter{equation}{0}
\setcounter{figure}{0}
\renewcommand{\theequation}{\Alph{section}.\arabic{equation}}
\renewcommand{\thefigure}{\Alph{section}\arabic{figure}}
It is insightful to examine the relations between the rates of change of quantities that appear in expression~(\ref{eq:model}).
For instance\,---\,for the case examined herein\,---\,what increase of speed would result from an increase of power by $1$~watt?

To answer this question, we need to find $\partial V_{\!\rightarrow}/\partial P$\,.
To do so\,---\,without solving equation~(\ref{eq:model}) for $V_{\!\rightarrow}$ as a function of~$P$, which is a cubic equation\,---\,we invoke the implicit function theorem.
Let us consider expression~(\ref{eq:misfit}). 
As required by the theorem, $f$ possesses continuous partial derivatives in all its variables at all points, except at $\lambda=1$\,, which is excluded by mechanical considerations.
Also, as required by the theorem, $f=0$\,, in the neighbourhood of interest, which is true as a consequence of equation~(\ref{eq:misfit}) and is illustrated in Figure~\ref{fig:Figfflat}.
Hence, in accordance with the theorem, among many relations between quantities in this expression, we can consider, for instance,
\begin{subequations}
	\begin{align}
		\dfrac{\,\,\,\partial V_{\!\rightarrow}}{\partial P}
		&=
		-\dfrac{
			\dfrac{\partial f}{\partial P}
		}{
			\dfrac{\partial f}{\partial V_{\!\rightarrow}}
		}
		\label{eq:ThmA}
		\\
		&=
		\dfrac{
			2\,(1-\lambda)
		}{
			2\,m\,a
			+
			\eta\,\rho\,{\rm C_{d}A}
			\left(V_{\!\rightarrow}+w_{\leftarrow}\right)
			\left(3\,V_{\!\rightarrow}+w_{\leftarrow}\right)
			+
			2\,m\,g
			\left({\rm C_{rr}}\cos\theta+\sin\theta\right)
		}
		\label{eq:ThmB}
		\,.
	\end{align} 	
\end{subequations}
Expression~(\ref{eq:ThmB}) is valid only in the neighbourhood of a point for which a combination of values\,---\,$P$\,, $\lambda$\,, $m$\,, $g$\,, $\theta$\,, $a$\,, ${\rm C_{rr}}$\,, $\eta$\,, ${\rm C_{d}A}$\,, $\rho$\,, $V_{\!\rightarrow}$\,, $w_{\leftarrow}$\,---\,results in~$f=0$\,.
It is not valid for arbitrary values of these quantities.
However, this restriction is not a significant limitation for a study of cycling performance, since $f=0$ is a criterion for an empirical adequacy of a power-meter model.

In accordance with Section~\ref{sec:NumEx}, inserting the fixed values, $m=111$\,, $g=9.81$\,, $w=0$\,,  which entails $\eta=1$\,, the averages of values obtained from measurements, $\overline P=258.8$ and $\overline V_{\!\rightarrow}=10.51$\,, and modelling, ${\rm C_{d}A}=0.2607$\,, ${\rm C_{rr}}=0.00231$\,, $\lambda=0.03574$\,, as well as letting $\overline\theta=0$\,, as an average over the entire segment, which is consistent with its flat topography, and $\overline a=0$\,, which is consistent with a steady tempo, we obtain the sought answer,
\begin{equation*}
	\dfrac{\partial V_{\!\rightarrow}}{\partial P}
	=
	0.0176846
	\,,
\end{equation*}
which means that an increase of power by $1$~watt results in an increase of speed of about $0.018$~metres per second.
Conversely, in accordance with expression~(\ref{eq:ThmA}), within this neighbourhood,
\begin{equation*}
	\dfrac{\partial P}{\partial V_{\!\rightarrow}}
	=
	56.5464
	\,,
\end{equation*}
which means that an increase of speed by $1$~metre per second requires an increase of power of about $57$~watts.
Thus\,---\,within the neighbourhood of~$f=0$\,---\,a $9.5$\% increase in speed requires about $22$\% increase in power.

\end{appendix}
\end{document}